# ANALYSIS OF SADDLE-POINT CONFIGURATIONS IN 3-DIMENSIONAL SU(2) LATTICE GAUGE THEORY

Chulwoo Jung [a] *

[a] Department of Physics, Columbia University, New York, NY 10027, USA

We discuss the properties of a class of saddle point solutions in SU(2) in three dimensions $(SU(2)_3)$, exhibiting localized peaks in the action. These configurations are generated by deterministic cooling and extremizing algorithms from analytic configurations. They share some characteristics with cooled and extremized Monte Carlo generated lattices. We have investigated physical behavior such as the string tension by averaging over this class of saddle point configurations. We have also measured the eigenvalues for harmonic fluctuations around these configurations.

## 1. Introduction

Confinement in non-Abelian gauge theories has been a very challenging problem to understand analytically. Even though there is considerable evidence for confinement at weak coupling from numerical simulations, the underlying property of gauge field configurations giving confinement has not been understood. This is largely due to the fact that confinement is a non-perturbative phenomenon, for which we have limited tools for investigation.

Cooling [1,2] (deterministic reduction of the action) or extremizing [3] (deterministic reduction of the Hessian, $\hat{S} = \sum_{i,x} \left( \partial S / \partial A^i(x) \right)^2$ ), when applied to Monte Carlo generated SU(2) lattices, tends to produce configurations with isolated peaks in the action density. And, remarkably, the string tension sustains its value after cooling has decreased the action to less than 1 percent of its initial value [1]. Since there are no stable nonzero solutions in $SU(2)_3$, the quasistability under cooling is likely due to being close to a saddle point solution with a few unstable modes.

Considering the fact that the saddle point can be defined precisely, in contrast to a "moderately cooled lattice", extremized lattices serve as a bet-

*This work was done in collaboration with Robert D. Mawhinney. It was partially supported by the US Department of Energy and the Pittsburgh Supercomputing Center.

ter staring point for quantitative investigation. Using these saddles, we could even try to get a valid approximation of the partition function by expanding around these saddle points:

$$\begin{aligned} Z &= \int dU_l e^{-S(U)} \\ &= \sum_i \int_{[i]} dU_l e^{-S(U)} \\ &\sim \sum_i e^{-S_{cl}(U_i)} \cdot D_i \end{aligned}$$

where $\int_{[i]} dU_l$ covers the gauge link configurations near a particular semiclassical solution, and $D_i$ is the fluctuation determinant near that solution, which is given by the product of the eigenvalues of harmonic fluctuations. (In the Georgi-Glashow model [4], the harmonic fluctuations were decoupled from the particular saddle, $D_i = $ constant.)

To achieve this, we need to (i) develop a systematic way to cover all the saddle points and (ii) get the eigenvalue spectrum around each saddle point. In this work, we use a starting configuration to make saddles which shares one characteristic common to many saddles found from Monte Carlo generated lattices. Section 2 describes the starting configuration and how we generate the saddle point solutions with this configuration. Section 3 discusses the result of string tension measurements of these saddle points. Section 4 discusses the eigenvalues of harmonic fluctuations around these configurations.

Figure 1. The starting configuration (Z(2) vortex) we use to produce saddle-point configurations. The monopole and anti-monopole (M and M̄) are defined relative to an arbitrary U(1) subgroup of SU(2). Wilson loops centered on the point P have a value of −1. Two such loops are shown by the dotted lines.

Figure 2. The action density plot of saddle-point configurations generated from a Z(2) vortex

## 2. Generation of saddle gas

On cooled lattices, we observed that Wilson loops centered on an action peak have trace very close to −1. This is the indication of Z(2) flux associated with the peak. So we have come up with the starting configuration in Fig. 1, where Wilson loops around P have trace = −1.

Fig 1 shows a monopole-antimonopole pair with respect to an arbitrary U(1) subgroup of SU(2). Flux from the monopoles is confined to the closed tube. Because this configuration is compact in size, we can distribute more than one of these vortices on a lattice without overlapping. Remarkably, most saddle point solutions obtained from more than one Z(2) vortex still exhibit well separated peaks after extremization. (We will call these saddle point solutions a " Z(2) saddle gas".) We can get various starting configurations by choosing different positions and orientations for the Z(2) vortices, and different U(1) subgroups. Starting from these configurations, we use deterministic cooling to get smooth configurations and then extremize further to get to the saddle point solutions. We used two different extremization algorithms, one by Duncan and Mawhinney [3] and the other by Van der Sijs [5]. The final saddle point solutions are quite independent of either the choice of extremization algorithm or the number of cooling steps initially taken.

## 3. String tension

Figure 3 shows the results of the measurement of the string tension of various SU(2) configurations. The string tension appears as a plateau in the Creutz ratio, given by

$$C(R,T) = -\ln\frac{W(R,T)W(R-1,T-1)}{W(R-1,T)W(R,T-1)}$$

where $W(R,T)$ is the average of the trace of a Wilson loop with size R×T.

As shown in the graph, the string tension sustains its value for cooled Monte Carlo lattices, but was apparently broken for extremized lattices [3].

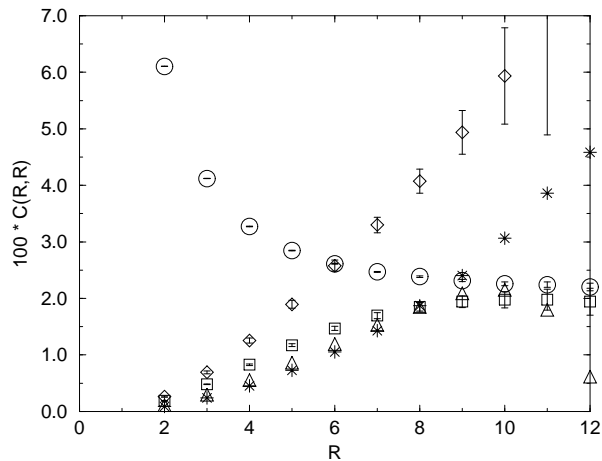

Figure 3. The Creutz ratios as a function of R for a Monte Carlo calculation at $\beta = 10.0$ ($\circ$), cooled Monte Carlo generated lattices ($\square$), extremized Monte Carlo generated lattices ($\diamond$), and two different 3-peak saddle-point lattices ($\triangle$ and $*$).

For Z(2) saddle gas configurations, we see two kinds of behavior: some configurations exhibit a rising Creutz ratio, while others have a Creutz ratio falling off at large distances. The fact that most of the configurations exhibit either a rising or falling of the Creutz ratio, not showing a plateau, shows that a plateau may be achieved by correct weighting.

## 4. Eigenvalues

The eigenvalues of harmonic fluctuations can be obtained by diagonalizing the matrix of the second derivatives of the action at the saddle point.

$$(A)_{ij} = \frac{\partial^2 S}{\partial A_i \partial A_j}$$

where $i, j$ are the indices for the gauge links.

Because the size of the matrix we need to diagonalize is large even for lattices of moderate size ($3 \times 3 \times N^3$ for $N^3$ lattice), we should use some indirect methods to calculate the eigenvalues. We use the Lanczos algorithm to tridiagonalize the matrix and use the bisection method to get the eigenvalues [6]. We have also implemented the spectral method [7] for comparison. For enough iterations, well separated eigenvalues agree between the two algorithms.

Table 1 shows the eigenvalues for a single and a double peak saddle. The first two columns are from a single peak saddle; the amount of extremization differs. The second two are from a configuration with two peaks. We didn't do any gauge fixing.

Highly degenerate eigenvalues, especially zero modes from the gauge degrees of freedom, pose some difficulty getting all the eigenvalues using these methods. In the first column of Table 1, we can see many small eigenvalues of $\mathcal{O}(10^{-4})$. These are actually zero eigenvalues, mainly from the gauge degrees of freedom, spread due to the fact that the configurations are not extremized enough. In the second column, the eigenvalues of $\mathcal{O}(10^{-4})$ have become $\mathcal{O}(10^{-10})$. This spreading happens to other eigenvalues as well, but is much more manifest for small eigenvalues. As you extremize further, the amount of spreading becomes smaller and finally becomes less than roundoff error, which is the case for the second column.

Because of these close eigenvalues, we needed more than 6×(size of the matrix) double precision Lanczos iterations to get reliable results for the lowest eigenvalues (whether we could get all the eigenvalues for less extremized lattices using enough iterations is not clear.) For the spectral method, many eigenvalues in such a small interval make it impractical to resolve all the eigenvalues. Using the spectral method, we would need to iterate for

$$\sim \frac{\lambda_{max} - \lambda_{min}}{\Delta \lambda}$$

steps, where $\lambda_{max}$ is the biggest eigenvalue, $\lambda_{min}$ is the smallest eigenvalue and $\Delta \lambda$ is the smallest difference between nearby eigenvalues. For the configuration in the first column, this is $\sim 10^8$ iterations.

As expected, these saddle point solutions have only a few negative eigenvalues, some of them degenerate, possibly from the symmetry of the starting configuration. The differences between eigenvalues of configurations with 1 peak and 2 peaks show that the quantum interaction with these Z(2) saddles can be important when we average over configurations.

| 1peak | 1peak | 2peak | 2peak |
| --- | --- | --- | --- |
| $\hat{S} = 4.29 \times 10^{-5}$ | $\hat{S} = 9.89 \times 10^{-12}$ | $\hat{S} = 4.56 \times 10^{-6}$ | $\hat{S} = 3.27 \times 10^{-9}$ |
| -3.4210669462e-01 | -3.4211895357e-01 | -5.2526713144e-01 | -5.2525807384e-01 |
| -1.1364556945e-01 | -1.1365255915e-01 | -4.8898471234e-01 | -4.8898469633e-01 |
| -1.1363992269e-01 | -7.7662707534e-02 | -1.8146869013e-01 | -1.8143397039e-01 |
| -7.7656950137e-02 | -5.5059548908e-10 | -1.8146088105e-01 | -1.8143395016e-01 |
| -7.7652752749e-02 | 1.9670999680e-01 | -1.8142878191e-01 | -1.5147668113e-01 |
| -7.7639221739e-02 | 3.0523180502e-01 | -1.5148221389e-01 | -1.3632465649e-01 |
| -6.8419454276e-04 | 3.4662400677e-01 | -1.5145419536e-01 | -1.3820967907e-06 |
| -5.3308552806e-04 | 3.5578869064e-01 | -1.5145150934e-01 | -1.3133195284e-06 |
| -4.7252056827e-04 | 3.7005482543e-01 | -1.3632724901e-01 | -1.1828021657e-06 |
| -4.1278359892e-04 | 3.9044932215e-01 | -1.3632386774e-01 | -1.1613608130e-06 |
| -3.5321133734e-04 | 4.1237209024e-01 | -1.7562655458e-04 | -8.9224434703e-07 |
| -3.4950329520e-04 | 4.1323358963e-01 | -1.7314719403e-04 | -7.8179389143e-07 |
| -3.2684186142e-04 | 4.2751186709e-01 | -1.6551398287e-04 | -7.2901517696e-07 |
| -2.9772940932e-04 | 5.7692296949e-01 | -1.5515659129e-04 | -7.0685911245e-07 |
| -2.9237841527e-04 | 5.8406633751e-01 | -1.5053838616e-04 | -6.6397640695e-07 |
| -2.7645002253e-04 | 6.5492639192e-01 | -1.2000722596e-04 | -6.4171038678e-07 |
| -2.6696879372e-04 | 6.7512327752e-01 | -1.0125968711e-04 | -6.0311595182e-07 |
| -2.5958680188e-04 | 7.4972946832e-01 | -9.6709858892e-05 | -5.7397770321e-07 |
| -2.5503394023e-04 | 7.5760661165e-01 | -9.4155278471e-05 | -5.4588403332e-07 |
| -2.4241598919e-04 | 7.6884680745e-01 | -9.2295679100e-05 | -5.0074723688e-07 |

Table 1
Lowest eigenvalues of Z(2) saddle gas configurations on a $12^3$ lattice from the Lanczos algorithm.

## 5. Discussion

Continuing study of the eigenvalue spectrum will show whether naive averaging of configurations will represent the partition function correctly. More eigenvalue measurements and comparison between the eigenvalues of the Z(2) saddle gas and Monte Carlo generated lattices are needed to study the validity of the starting configurations we have used. Also, measuring other physical quantities, such as the chiral condensate, on these configurations will give us an interesting perspective about the nature of confinement.

## REFERENCES

1. M. Campostrini, A. Di Giacomo, M. Maggiore, H. Panagopoulos and E. Vicari, *Phys. Lett.* **B225** (1989) 403.
2. A. Duncan and R. Mawhinney, *Phys. Lett.* **B241** (1990) 403, *Phys. Rev.* **D43** (1991) 554.
3. A. Duncan and R. Mawhinney, *Phys. Lett.* **B282** (1992) 423.
4. A.M. Polyakov, *Nucl. Phys.* **B120** (1977) 429.
5. A.J. van der Sijs, *Nucl. Phys.* **B (Proc. Suppl.) 30** (1993) 893, *Phys. Lett.* **B294** (1992) 391.
6. I.M. Barbour, N.E. et al. DESY 84/087(1984) Published in *Lecture Notes in Physics, The Recursion Method and its Applications*, Springer-Verlag (1985) 149.
7. C. Lanczos, *Applied Analysis*, Dover (1988) 180.